\documentstyle[twocolumn,aps]{revtex}
\begin{document}
\draft

\title{{\it In-Situ} Infrared Transmission Study of Rb- and K-Doped Fullerenes}

\author{Michael C. Martin, Daniel Koller, L. Mihaly}

\address{Department of Physics, State University of New York at Stony Brook,
Stony Brook, NY 11794-3800}

\date{23 December 1992}
\maketitle

\begin{abstract}
We have measured the four IR active $C_{60}$ molecular vibrations in
$M_{x}C_{60}$ $(M = K, Rb)$ as a function of doping $x$. We observe
discontinuous changes in the vibrational spectra showing four distinct
phases (presumably $x = 0, 3, 4$, and 6).  The $1427cm^{-1}$ and
$576cm^{-1}$ modes show the largest changes shifting downward in
frequency in four steps as the doping increases.  Several new very
weak modes are visible in the $x=6$ phase and are possibly Raman modes
becoming weakly optically active.  We present quantitative fits of the
data and calculate the electron-phonon coupling of the $1427cm^{-1}$ IR mode.
\end{abstract}
\narrowtext

The discovery of superconductivity in alkali doped $C_{60}$
\cite{discsuper} has prompted a great deal of excitement and a large body
of research.  Since
electron-phonon coupling proved to be responsible for superconductivity
in many materials, the
relation between the charge carriers and the lattice vibrations
is important.  Here we present a study of the
infrared active molecular vibrations of $M_{x}C_{60}$ as a function
of alkali metal doping,$x$, for $(M = K$ and $Rb)$.
We show, for the first time, the shift of some IR modes to lower
frequencies in discrete steps, corresponding to the four
known stable phases, $x=0,3,4,6$.
We also perform an analysis of how the change in the vibrational modes
is related to the introduction of carriers into the lowest unoccupied
molecular orbital band of $C_{60}$.

The truncated icosohedral structure of $C_{60}$ fullerenes belongs to the
icosohedral point group, $I_{h}$, and has four infrared active intramolecular
vibrational modes with $F_{1u}$ symmetry \cite{modestheory}.  These modes,
with center frequencies $\nu _{1}=527,
\nu _{2}=576, \nu _{3}=1182$, and $\nu _{4}=1427$ have been
experimentally observed
\cite{earlyIR,Kratschmer}.
The 527 and $576 cm^{-1}$ modes are associated with primarily radial motion of
the carbon atoms while the 1182 and $1427 cm^{-1}$ modes are primarily
tangential motion \cite{modestheory}.

In the $M_{x}C_{60}$ compounds, the
alkali atoms give up one electron each to the lowest unoccupied molecular
orbital (LUMO) of a $C_{60}$ molecule.
As long as the on site Coulomb repulsion is not too large
the triply degenerate $t_{1u}$ LUMO can hold six electrons
\cite{huckel}.  Therefore the $t_{1u}$ orbital is half filled by three
electrons, and the material is metallic
and superconducts at low temperatures\cite{discsuper}.  Completely
filling the $t_{1u}$ orbital with six electrons makes the $C_{60}$ a band
insulator and the structure becomes body-centered-cubic \cite{bccx=6}.
An insulating phase at $x=4$ has also been observed
with a body-centered-tetragonal structure \cite{bctx=4}.

The $C_{60}$ for this study was prepared using the well known
technique of Kratschmer {\it et. al.} \cite{Kratschmer}.  The resultant
fullerene powder was loaded into a tantalum boat and heated to
about $500^{o}C$ in a vacuum of $\sim 1$x$10^{-6}$ Torr to vapor deposit
$C_{60}$ inside the sample cell.  Film thickness was monitored by counting
visible light interference fringes as the $C_{60}$ was deposited.
Typical sample thickness was about $1.2\mu m$.

  Our miniature sample cells are constructed of glass with two silicon
windows, one of which serves as a substrate for the $C_{60}$.
A small appendage contains the alkali metal and the entire
sample chamber is sealed under high vacuum.   The transmission spectra
were obtained with a Bomem $MB-155$ FTIR spectrometer
at $2cm^{-1}$ resolution covering a $400 - 6000 cm^{-1}$ frequency range
as the sample was doped.  Four-probe resistivity of the film was
measured simultaneously.  By carefully warming the
alkali metal to increase the vapor pressure and by heating the $C_{60}$
film to increase the diffusion rate of the metal into the film, a slow,
continuous doping was achieved.  During potassium doping, the substrate
was maintained at $120^{o}C$ and the metal at $100^{o}C$.  For $Rb$, the
corresponding temperatures were $85^{o}C$ and $70^{o}C$ respectively.  The
infrared spectra were obtained with the film samples at their substrate
temperatures.

The results of the infrared spectroscopy {\it in-situ} with alkali
metal doping are displayed in Figures \ref{Rbdoping} and \ref{Kdoping}.
Since at low frequencies the transmitted intensity is roughly proportional
to the square of the resistivity, the IR measurement by itself is a good
indicator of the doping process.  Indeed, both the DC resistivity and the
transmission exhibit a minimum as the doping proceeds.  Kochanski {\it et.
al.} associate this minimum with the metallic $M_{3}C_{60}$ phase
\cite{Vee-shape}.  Upon further doping, the resistivity reaches a maximum,
corresponding to the insulating $M_{6}C_{60}$ phase, after which the
resistivity drops, but the features in the spectra change little,
indicating that we are now just coating the sample with
alkali metal and the $C_{60}$ is fully doped $(x=6)$.
At this stage, closer visual inspection reveals metallic films on the
window surfaces.

We have performed quantitative fits to both sets of doping data.  The
vibrational spectra are accuratly fit using a dielectric function composed
of Lorentzian oscillators and a dc conductivity to mimic the changing
conductivity observed as doping progresses. This yields the center
frequency, $\omega _{0}$, strength, S, and width, $\Gamma $, of each mode.  We
present the results and our assignments of corresponding phases in Table
\ref{dopingtable}.  We interpret the data as follows:  as the doping proceeds,
layers of $M_{3}C_{60}$ grow followed by similar layers of $M_{4}C_{60}$ and
$M_{6}C_{60}$.  We observe that the $\nu _{1}$ mode at $526cm^{-1}$
gets weaker as the sample looses all $x=0$
phase and reappears at $468cm^{-1}$ in the $x=6$ phase.  The $\nu _{2}$ mode
shifts in
discrete steps and grows in strength during the doping process.  This is
most clearly visible for the $Rb$ doping where we see it shifts from 576 to
573 to 570 to $565cm^{-1}$.  Using the $x$-ray results
\cite{fccx=0,fccx=3,bctx=4,bccx=6} which indicate that only
the $x=0, 3, 4$, and 6 phases are stable at this temperature, we associate
these modes with the $x = 0, 3, 4$, and 6 phases respectively.  The
$\nu _{3}$ mode at $1182cm^{-1}$ is enhanced by a factor of almost 2 in
the $x=6$ phase.  Looking at the $\nu _{4}$ mode, we can again see
distinct phases.  This mode is enhanced by a factor of 80 in the $x=6$ phase.

The shifts in the $\nu _{2}$ and $\nu _{4}$ modes for the K doped sample show
only three clear phases.  This may be due to more uniform doping of the
$K_{x}C_{60}$ sample, as it is known that K diffuses more rapidly into
$C_{60}$ than $Rb$ \cite{raman}.
Vibrational modes are more difficult to detect in a homogeneous metallic
film since the high reflectivity, due to conduction electrons, dominates
the response.

The resultant values for our $x=6$ data are in excellent agreement
with previous IR measurements on $M_{6}C_{60}$ published by Fu {\it et. al.}
\cite{fu}.
The phase separation seen in our data is also in agreement with previous
Raman spectroscopy results \cite{raman,raman1} where, for example, the
change in the $A_{g}(2)$ pinch mode originally at $1458cm^{-1}$ clearly shows
distinct phase separation.  Note that this mode, like the $\nu_{4}$ mode
in the IR spectrum, involves stretching of the $C=C$ double bonds.

The $Rb_{6}C_{60}$ and $K_{6}C_{60}$ spectra also show some very weak modes
at 1461, 1418, 1317, 1284, 1240, 1190, 1146, 943, 688, 645 and
$532cm^{-1}$.  These are similar in frequency to several Raman active
modes for $C_{60}$ (1458, 1430, 1396, 1315, 1241, 1190,
1140, 950, and $533cm^{-1}$) and $Rb_{6}C_{60}$ (1477, 1431, 1322,
1235, 1121, 1091, 689, and $655cm^{-1}$)\cite{raman,notallowed,mitch}.  We may
be observing Raman modes becoming weakly IR-active in our fully doped
samples.  Some of these modes are not allowed by the symetry of the
single molecule, but they have been observed in Raman spectra of crystaline
$C_{60}$ \cite{notallowed,mitch}.
It should be noted that the changes in the vibrational modes due to
photoexcitation of carriers \cite{photoinduced} are different from the changes
reported here for chemical doping.

The charged-phonon theory by M. J. Rice and Han-Yong Choi \cite{rice} predicts
many of the features we observe, in particular, the enhancement of the
strength and the softening of the $1428cm^{-1}$ mode upon doping.
In Figure \ref{molecule} we illustrate how the addition of electrons to
a complex molecule can enhance the strength of a "silent" IR active phonon.
Figure \ref{molecule} (a) depicts an eigenmode of the molecule,
characterized by two pairs of atoms oscillating in opposite phase.
This mode is
"IR active" in the sense that it has odd symmetry.  However, as long as only
symetric electronic states are allowed, the mode is "silent" since there is
no net dipole moment in the direction of the electric field.
In Figure \ref{molecule} (b) we allow for asymmetric states, i.e.
electon transfer between
the two pairs of atoms.  The electron transfer naturally couples to the
atomic displacements via the rearrangement of electronic states, and it also
generates an electric dipole moment.  Rice and Choi
\cite{rice} argue that in $C_{60}$, electrons excited between the
between the $t_{1u}$ and the next higher energy $t_{1g}$ molecular orbitals
create a coupling to the infrared active vibrational modes, and
simultaneously soften the vibration frequency (Figure \ref{agreement}).
In view of the agreement between the experiment and theory for the
$\nu_{4}$ mode, it is all the more surprising that the $\nu_{1}$ mode
does not follow the same behavior.  We argue that this mostly radial
mode is more sensitive to the inter-molecular forces, and its frequency
is influenced by the FCC to BCC phase transition.

As long as the electron is confined to a single molecule, symmetry arguments
predict that the intraband electron-phonon scattering, relevant to
superconductivity, is dominated by Raman modes.
Therefore, the attempts to explain superconductivity in the fullerenes
focussed on the Raman active $A_{g}$ and $H_{g}$ modes \cite{el-phtheory}.
However, for extended electronic or vibrational states, the IR modes may
become important too.  In fact, we see evidence for damping of the
vibrational modes by conduction electrons.
According to the sum rule,
$\sum_{i} \Gamma _{i}/\omega _{0i}^{2} = (\pi /2)N(E_{F})\lambda$ \cite{pba},
the broadening of the vibrational mode is related to the electron-phonon
coupling. The largest change, $16.3cm^{-1}$, in the widths was observed in
the $\nu_{4}$ mode.  This
leads to a contribution of 0.055 states/eV in $N(E_{F})\lambda$,
for each of the three degenerate modes.  Considering that there are 180
vibrational modes for the $C_{60}$ molecule, a coupling of this magnitude
is significant although the strongest electron-phonon coupling seen to date
is for the $H_{g}(2)$ Raman mode (0.25 states/eV) \cite{mitch}.
It remains to be seen how important this coupling is for superconductivity.

In summary, we have measured the four infrared active intramolecular
vibrational modes of $C_{60}$ as a function of $Rb$ and $K$ doping.
Quantitative presentation of our measurements and assignments of the modes to
different stable phases of $M_{x}C_{60}$ were made.  We show that our results
are consistent with previous works and describe how our data could relate to
a superconducting pairing mechanism in these materials.

\acknowledgments

This work was supported by NSF grant DMR9202528.  We would like to
thank P. B. Allen,G. L. Carr and K. Holczer for valuable discussions.  Thanks
to J. Marecek of the SUNY at Stony Brook chemistry department for $C_{60}$
separation and purification and to J. Kirz and S. Lindaas for the use of high
vacuum equipment.

\begin{figure}
\caption{$Rb_{x}C_{60}$ infrared transmission spectra as a function of x.  (a)
shows the lower two vibrational modes $(\nu _{1}$ and $\nu _{2})$ and
(b) shows the upper two modes $(\nu _{3}$ and $\nu _{4})$.  The curves
are offset for clarity.  Undoped $C_{60}$ is the top curve and the sample is
further doped going down in the figure.  The bottom curve is for fully
doped $Rb_{6}C_{60}$.  The minimum in the resistivity of the sample corresponds
to the curve labeled 'min'.  Vertical lines are guides to indicate the
assignments of vibration frequencies, as obtained from numerical fits.}
\label{Rbdoping}
\end{figure}

\begin{figure}
\caption{The same as Figure 1 now for $K_{x}C_{60}$.}
\label{Kdoping}
\end{figure}

\begin{figure}
\caption{Schematic illustration of charge-transfer induced infrared activity.
The arrowed circles represent atoms moving withing a single "molecule";
the symmetry of the oscillation is odd.
In (a) only even electronic states are allowed and therefore the
strength of the IR active vibration is small.  In (b), the electron
transfer couples the mode to the external field by introducing a
dipole moment.}
\label{molecule}
\end{figure}

\begin{figure}
\caption{The frequency of the $\nu_{4}$ IR mode vs. the inferred composition
for $Rb$ doped $C_{60}$.  Circles refer to the calculation of
Rice and Choi (Ref. [15]). }
\label{agreement}
\end{figure}

\begin{table}
\caption{Lorentzian oscillator fits to the four IR intramolecular
vibrations of $M_{x}C_{60}$ $(M=Rb$ and $K)$ assigned to the different stable
phases $(x=0, 3, 4, 6)$.  The numbers are essentially the same for both dopants
except ($^{a}$) as noted.  $\omega _{0}$ is the center frequency, $S$ is the
strength $(S=\omega _{p}^{2}/\omega _{0}^{2})$, and $\Gamma $ is the width of
each vibrational mode.}
\begin{tabular}{cccccc}
&&$C_{60}$&$M_{3}C_{60}$&$M_{4}C_{60}$&$M_{6}C_{60}$\\
\tableline
&$\omega _{0}(cm^{-1})$&526&&472&467\\
$\nu _{1}$&$S$&.02&&.008&.03\\
&$\Gamma  (cm^{-1})$&2.5&&1.5&3\\
\tableline
&$\omega _{0}(cm^{-1})$&576&573&570&565\\
$\nu _{2}$&$S$&.008&.019&.022&.17\\
&$\Gamma  (cm^{-1})$&2.7&3&3.7&2.8\\
\tableline
&$\omega _{0}(cm^{-1})$&1182&&&1182\\
$\nu _{3}$&$S$&.0018&&&.003\\
&$\Gamma  (cm^{-1})$&4.2&&&5.8\\
\tableline
&$\omega _{0}(cm^{-1})$&1428&1393&$1363\tablenote{Values in table
are for $Rb_{4}C_{60}$.  $K_{4}C_{60}$ differs for $\nu _{4}: \omega _{0}=
1369cm^{-1}, S = .028$, and $\Gamma  = 21cm^{-1}$.}$&1340\\
$\nu _{4}$&$S$&.001&.012&.016&.08\\
&$\Gamma  (cm^{-1})$&4.5&20.8&23&7.2\\
\end{tabular}
\label{dopingtable}
\end{table}

\end{document}